\documentclass[aps,prd]{revtex4-2}

\usepackage[T1]{fontenc}
\usepackage[utf8]{inputenc}

\usepackage{mathptmx}
\usepackage{amsmath}
\usepackage{amssymb}
\usepackage{upgreek}
\usepackage{moresize}
\usepackage{microtype}
\usepackage{graphicx}

\usepackage{color}
\usepackage{soul}

\begin{document}

\title{Scaling properties of nuclear parton distributions in short-range-correlation motivated two-component parametrization}

\author{Petja Paakkinen}
\email{petja.k.m.paakkinen@jyu.fi}
\affiliation{University of Jyväskylä, Department of Physics, P.O. Box 35, FI-40014 University of Jyväskylä, Finland\\ and Helsinki Institute of Physics, P.O. Box 64, FI-00014 University of Helsinki, Finland}

\begin{abstract}
We provide some critical remarks on the recently proposed two-component parametrization of nuclear parton distribution functions, which was motivated by the apparent correlation between the nuclear modifications of structure functions and nucleon-nucleon short-range correlation phenomena. This parametrization, we show, is invariant under a rescaling transformation of the involved abundance coefficients, which means that the global normalization of these coefficients cannot be meaningfully determined in a fit, and only their ratios should be studied for finding evidence of short-range-correlation type behavior at parton level. As we show, however, the current constraints for the nuclear-mass dependence of these coefficients allow also for interpretations different from short-range correlations. Nevertheless, this two-component parametrization exhibits a similar scaling relation for DIS structure functions as demonstrated in earlier works, and, as we demonstrate, yields testable predictions for structure-function and hard-process cross-section ratios. We also note on the non-trivial isospin dependence of the short-range-correlation motivated parametrization, which under proton-neutron pair dominance assumption can lead to charge-symmetry-violation resembling terms.
\end{abstract}

\maketitle
\vspace{-0.5cm}

\section{Introduction}

The mechanism behind nuclear modifications of nucleon partonic structure has been subject to immense discussion ever since their first observation in early fixed-target lepton-nucleus deep inelastic scattering (DIS) experiments~\cite{Arneodo:1992wf}. Very recently, a new way of connecting these parton-level nuclear modifications with the nucleonic nuclear structure was proposed in terms of a short-range correlation (SRC) motivated parametrization of the nuclear parton distribution functions (nPDFs)~\cite{PhysRevLett.133.152502}. This idea was based on earlier evidence of a correlation between the DIS structure function and quasielastic (QE) scattering data, which has been linked with nucleon-nucleon SRCs~\cite{RevModPhys.89.045002}. The new approach relies on two assumptions: the partonic structure of the nucleons in nuclei are unmodified compared to free nucleons, except when they are bound in SRC pairs, and the modifications in SRC pairs are universal, independent of the nucleus. Within this two-component model, one then parameterizes the PDFs for nucleons in SRC pairs, fitting the parameters with global hard-process cross-section data, while taking the unmodified quasi-free nucleon PDFs as fixed from a separate global analysis. This contrasts the new approach from conventional global fits of nPDFs~\cite{annurev-nucl-102122-022747}, where one traditionally parameterizes the (effective) bound-nucleon PDFs without separating them into two distinct components.

This article should be seen as both appraisal and criticism of the approach taken in Ref.~\cite{PhysRevLett.133.152502}. First of all, we credit the work in Ref.~\cite{PhysRevLett.133.152502} for taking the SRC-motivated approach to the parton-distribution global-analysis level, which makes it possible to extend the analysis beyond DIS structure functions where the approach is typically applied in, and, as we will show, leads to predictions that can be potentially tested against future data from the Electron-Ion Collider (EIC) and CERN Large Hadron Collider (LHC). In particular, we find scaling relations for ratios of structure functions and cross sections of different-mass nuclei, similarly to what has been discussed in Refs.~\cite{CLAS:2019vsb,PhysRevLett.124.092002,Xu:2019wso,PhysRevD.104.033002,PhysRevD.108.053005,Huang:2025kmd} but noting that their validity extends to a wide class of observables and for isoscalar nuclei also beyond the proton-neutron pair dominance assumption. However, we also provide some, hopefully constructive, criticisms by pointing out that the way the SRC-motivated parametrization is done in Ref.~\cite{PhysRevLett.133.152502} suffers from a rescaling invariance which makes the normalization of the modified-nucleon abundance coefficients extracted there non-unique and specific to the applied parametrization of the SRC PDFs. As we will show, only the ratios of these coefficients can be constrained meaningfully in an nPDF fit, for which purpose we provide an alternative formulation of the SRC-motivated parametrization. We further argue that these coefficients allow for other interpretations than being SRC abundances, and the current constraints on their nuclear-mass dependence can be described by any model with an approximate volume scaling with a surface correction. In addition, we show that assuming isospin symmetry for the SRC PDFs and proton-neutron pair dominance for the abundance coefficients leads to charge-symmetry-violation (CSV) like terms for the average bound-nucleon PDFs.

\section{Nuclear PDFs from SRC PDFs in two-component approach}

It is customary to decompose the nPDFs as
\begin{equation}
  f^A_i(x,Q) = \frac{Z}{A} f^{p/A}_i(x,Q) + \frac{N}{A} f^{n/A}_i(x,Q),
  \label{eq:full-nucleus-from-bound-nucleons}
\end{equation}
where $f^A_i(x,Q)$ is the PDF of the full nucleus with $Z$ protons and $N = A - Z$ neutrons (normalized to per-nucleon level), and $f^{p/A}_i(x,Q)$, $f^{n/A}_i(x,Q)$ denote the bound-proton and bound-neutron PDFs, respectively. Here, $x$ is the (per-nucleon) momentum fraction and $Q$ the factorisation scale. As has been stressed many times in the literature, these bound-nucleon PDFs are an effective description of how the parton distributions of the protons and neutrons bound in the nucleus look \emph{on average}. Eq.~\eqref{eq:full-nucleus-from-bound-nucleons} should thus not be interpreted in the way as indicating each and every bound proton and neutron being modified the same way, but works as a more general, averaged, description of the nPDFs.

In the SRC-motivated nPDF parametrization of Ref.~\cite{PhysRevLett.133.152502}, these \emph{effective} average bound-nucleon PDFs are written as
\begin{equation}
  \begin{split}
    f^{p/A}_i(x,Q) &= (1-C^A_p) \times f^p_i(x,Q) + C^A_p \times f^{\mathrm{SRC}p}_i(x,Q), \\
    f^{n/A}_i(x,Q) &= (1-C^A_n) \times f^n_i(x,Q) + C^A_n \times f^{\mathrm{SRC}n}_i(x,Q),
  \end{split}
  \label{eq:bound-nucleon-from-src}
\end{equation}
separating the contributions from the unmodified quasi-free nucleons $f^p_i(x,Q)$, $f^n_i(x,Q)$, and from the nucleons in SRC pairs $f^{\mathrm{SRC}p}_i(x,Q)$, $f^{\mathrm{SRC}n}_i(x,Q)$. The relative size of the contribution from the latter on the full nucleus PDFs is determined by the SRC-nucleon abundance coefficients $C^A_p$, $C^A_n$. Similar SRC-motivated decompositions have appeared previously for DIS structure functions~\cite{CLAS:2019vsb,PhysRevLett.124.092002} and at the level of PDFs~\cite{Xu:2019wso,PhysRevD.104.033002,PhysRevD.108.053005,Huang:2025kmd}, typically within the proton-neutron ($pn$) pair dominance assumption where the number of modified pairs is fixed to $n^A_{pn} = Z \, C^A_p = N \, C^A_n$, but Ref.~\cite{PhysRevLett.133.152502} appears to be the first publication applying this formalism to a global fit of nPDFs in a simultaneous analysis of multiple observables.

It should be noted that Eq.~\eqref{eq:bound-nucleon-from-src} is quite general, and nuclear effects arising from phenomena different to nucleon SRCs could manifest themselves in a similar form. In fact, ``SRC'' could be traded here with any ``exotic'' component, cf.\ Ref.~\cite{PhysRevLett.52.2344}. Still, as will be shown below, Eq.~\eqref{eq:bound-nucleon-from-src} is more restrictive than a fully general form of the bound-nucleon PDFs $f^{p/A}_i(x,Q)$, $f^{n/A}_i(x,Q)$, and leads to testable predictions. Note, in particular, that due to the linearity of Dokshitzer-Gribov-Lipatov-Altarelli-Parisi (DGLAP) equations~\cite{Dokshitzer:1977sg,Gribov:1972ri,Lipatov:1974qm,Altarelli:1977zs}, if Eqs.~\eqref{eq:full-nucleus-from-bound-nucleons} and~\eqref{eq:bound-nucleon-from-src} hold at a particular scale $Q$, they must hold also at any other perturbative scale.

\section{Isospin relations}

In the conventional global nPDF fits, one typically assumes the isospin symmetry (IS) between the average bound-nucleon PDFs, $f^{p/A}_u(x,Q) = f^{n/A}_d(x,Q)$, $f^{p/A}_d(x,Q) = f^{n/A}_u(x,Q)$, etc. This in fact implicitly assumes that these bound-nucleon PDFs depend on the nucleus only through the mass number $A$, independently of the $Z/N$ ratio. With the $C^A_p$, $C^A_n$ coefficients in the SRC motivated approach being in principle allowed to differ for isobar nuclei, we have to expand the notation to find the correct isospin relations. In general terms, swapping $u$ and $d$ quarks under the IS transforms a state with $Z$ protons and $N$ neutrons into a state with $Z' = N$ protons and $N' = Z$ neutrons, which under the operator definition of parton distributions~\cite{Collins:2011zzd} requires that
\begin{equation}
  f^{(Z,N)}_u(x,Q) = f^{(Z'=N,N'=Z)}_d(x,Q), \qquad f^{(Z,N)}_{\bar{u}}(x,Q) = f^{(Z'=N,N'=Z)}_{\bar{d}}(x,Q), \qquad f^{(Z,N)}_{i \neq u, d, \bar{u}, \bar{d}}(x,Q) = f^{(Z'=N,N'=Z)}_i(x,Q).
  \label{eq:mirror-nuclei-is-relations}
\end{equation}
As special cases, for isoscalar nuclei ($Z = N = A/2$) one has
\begin{equation}
  f^{(Z=N)}_u(x,Q) = f^{(Z=N)}_d(x,Q), \qquad f^{(Z=N)}_{\bar{u}}(x,Q) = f^{(Z=N)}_{\bar{d}}(x,Q),
\end{equation}
and for free nucleons $p = (Z=1,N=0)$, $n = (Z=0,N=1)$ the typical relations
\begin{equation}
  f^p_u(x,Q) = f^n_d(x,Q), \qquad f^p_d(x,Q) = f^n_u(x,Q), \qquad f^p_{\bar{u}}(x,Q) = f^n_{\bar{d}}(x,Q), \qquad f^p_{\bar{d}}(x,Q) = f^n_{\bar{u}}(x,Q), \qquad f^p_{i \neq u, d, \bar{u}, \bar{d}}(x,Q) = f^n_i(x,Q)
  \label{eq:free-nucleon-is-relations}
\end{equation}
are recovered.

These relations do not fix the isospin relations of the SRC PDFs or the associated abundance coefficients completely, but additional assumptions are needed also in this case. By extending the free-nucleon IS relations to hold also for the quasi-free nucleons and similarly for the SRC PDFs,
\begin{equation}
  \begin{split}
    f^{\mathrm{SRC}p}_u(x,Q) &= f^{\mathrm{SRC}n}_d(x,Q), \qquad f^{\mathrm{SRC}p}_d(x,Q) = f^{\mathrm{SRC}n}_u(x,Q), \\ f^{\mathrm{SRC}p}_{\bar{u}}(x,Q) &= f^{\mathrm{SRC}n}_{\bar{d}}(x,Q), \qquad f^{\mathrm{SRC}p}_{\bar{d}}(x,Q) = f^{\mathrm{SRC}n}_{\bar{u}}(x,Q), \qquad f^p_{i \neq u, d, \bar{u}, \bar{d}}(x,Q) = f^n_i(x,Q),
  \label{eq:src-nucleon-is-relations}
  \end{split}
\end{equation}
it follows from Eq.~\eqref{eq:mirror-nuclei-is-relations} that
\begin{equation}
  \frac{Z}{A} \, \left(C^{(Z,N)}_p - C^{(Z'=N,N'=Z)}_n\right) \, [f^{\mathrm{SRC}p}_u(x,Q) - f^p_u(x,Q)] + \frac{N}{A} \, \left(C^{(Z,N)}_n - C^{(Z'=N,N'=Z)}_p\right) \, [f^{\mathrm{SRC}p}_d(x,Q) - f^p_d(x,Q)] = 0,
\end{equation}
with similar relations for $i \neq u, d$. Assuming non-zero SRC PDF modifications, $f^{\mathrm{SRC}p}_i(x,Q) - f^p_i(x,Q) \neq 0$, that are universal (independent of $A$) but flavor dependent, the above can hold for all nuclei simultaneously only if
\begin{equation}
  C^{(Z,N)}_p = C^{(Z'=N,N'=Z)}_n
  \label{eq:mirror-nuclei-coeff-is-relation}
\end{equation}
applies for mirror nuclei. For isoscalar nuclei, the above assumptions imply that
\begin{equation}
  C^{(Z=N)}_p = C^{(Z=N)}_n.
  \label{eq:isoscalar-coeff-is-relation}
\end{equation}
For the special case of perfectly $u$-$d$ symmetric SRC modifications, $f^{\mathrm{SRC}p}_u(x,Q) - f^p_u(x,Q) = f^{\mathrm{SRC}p}_d(x,Q) - f^p_d(x,Q)$ ($\bar{u}$, $\bar{d}$ similarly), one has instead a more loose condition that $\frac{Z}{A} \, (C^{(Z,N)}_p - C^{(Z'=N,N'=Z)}_n) + \frac{N}{A} \, (C^{(Z,N)}_n - C^{(Z'=N,N'=Z)}_p) = 0$, which is fulfilled automatically for isoscalar nuclei without the need of imposing Eq.~\eqref{eq:isoscalar-coeff-is-relation}. However, only the sum $C^{(Z=N)}_p + C^{(Z=N)}_n$ would enter in any physical predictions in this case, leaving their difference irrelevant. Also, if the $pn$ pair dominance is assumed, Eq.~\eqref{eq:isoscalar-coeff-is-relation} follows by construction. We will thus assume that Eq.~\eqref{eq:isoscalar-coeff-is-relation} holds, as imposed also in Ref.~\cite{PhysRevLett.133.152502}.

Furthermore, applying Eqs.~\eqref{eq:free-nucleon-is-relations} and~\eqref{eq:src-nucleon-is-relations} to the definition of the average bound nucleon PDFs in Eq.~\eqref{eq:bound-nucleon-from-src}, we find
\begin{equation}
  \begin{split}
    f^{n/A}_u(x,Q) &= f^{p/A}_d(x,Q) - (C^A_p - C^A_n) \, [f^{\mathrm{SRC}p}_d(x,Q) - f^p_d(x,Q)], \\
    f^{n/A}_d(x,Q) &= f^{p/A}_u(x,Q) - (C^A_p - C^A_n) \, [f^{\mathrm{SRC}p}_u(x,Q) - f^p_u(x,Q)].
  \end{split}
  \label{eq:bound-nucleon-is-relations-from-src}
\end{equation}
That is, if IS is assumed to hold for the SRC PDFs as in Eq.~\eqref{eq:src-nucleon-is-relations}, this produces for nuclei with $C^A_p \neq C^A_n$ terms resembling those of CSV~\cite{RevModPhys.82.2009}, similarly to the case of an isovector mean field PDF modification~\cite{PhysRevLett.102.252301}. However, as stressed in Refs.~\cite{RevModPhys.82.2009,PhysRevLett.102.252301}, these CSV-\emph{like} terms should not be interpreted directly as CSV in the electroweak-physics sense since they can originate from genuine nuclear effects, and comply with the global IS. If, conversely, $f^{n/A}_u(x,Q) = f^{p/A}_d(x,Q)$ is forced at the bound-nucleon level, this necessarily requires for non-vanishing SRC-nucleon modifications that $C^A_p = C^A_n$ for all nuclei. We thus see that assuming the proton-neutron pair dominance $Z \, C^A_p = N \, C^A_n$ necessarily leads to CSV-like terms for non-isoscalar nuclei at the level of bound-nucleon PDFs, if Eq.~\eqref{eq:src-nucleon-is-relations} is assumed. We should note, however, that assuming Eq.~\eqref{eq:src-nucleon-is-relations} is not strictly necessary, but it does reduce the degrees of freedom that need to be fitted significantly, similarly to the bound-nucleon IS assumption in conventional nPDF fits.

\section{Rescaling invariance}

We note that the nPDFs defined in terms of Eq.~\eqref{eq:bound-nucleon-from-src} are invariant under the following transformation:
\begin{equation}
  \begin{split}
    C^A_p &\to \tilde{C}^A_p = X \, C^A_p, \qquad
    f^{\mathrm{SRC}p}_i(x,Q) \to \tilde{f}^{\mathrm{SRC}p}_i(x,Q) = X^{-1} \, f^{\mathrm{SRC}p}_i(x,Q) + (1 - X^{-1}) \, f^p_i(x,Q), \\
    C^A_n &\to \tilde{C}^A_n = Y \, C^A_n, \qquad
    f^{\mathrm{SRC}n}_i(x,Q) \to \tilde{f}^{\mathrm{SRC}n}_i(x,Q) = Y^{-1} \, f^{\mathrm{SRC}n}_i(x,Q) + (1 - Y^{-1}) \, f^n_i(x,Q),
  \end{split}
  \label{eq:src-rescaling-trasformation}
\end{equation}
where $X$ and $Y$ are arbitrary nonzero scalar factors. The reason behind this invariance can be understood by rewriting Eq.~\eqref{eq:bound-nucleon-from-src} as
\begin{equation}
  \begin{split}
    f^{p/A}_i(x,Q) &= f^p_i(x,Q) + C^A_p \, [f^{\mathrm{SRC}p}_i(x,Q) - f^p_i(x,Q)], \\
    f^{n/A}_i(x,Q) &= f^n_i(x,Q) + C^A_n \, [f^{\mathrm{SRC}n}_i(x,Q) - f^n_i(x,Q)].
  \end{split}
  \label{eq:bound-nucleon-from-src-pdf-difference}
\end{equation}
Here, one can readily observe that any downward or upward rescaling of the coefficients $C^A_p$, $C^A_n$ can be compensated by adjusting the SRC PDFs to make their difference with the free-nucleon PDFs respectively bigger or smaller. We stress that the transformation in Eq.~\eqref{eq:src-rescaling-trasformation} can be done for any fixed free-nucleon PDFs, and it leaves the full-nucleus invariant, $f^A_i(x,Q) \to f^A_i(x,Q)$. Therefore, all physical observables, and any PDF sum rules, are preserved in the transformation.

Since the SRC PDFs are taken as universal, independent of the nucleus, this rescaling is a global one, i.e.\ all the $C^A_p$, $C^A_n$ coefficients must be scaled simultaneously with the same factors $X$ and $Y$, respectively, for all $A$. Furthermore, if one assumes the isospin symmetry for the SRC PDFs as given in Eq.~\eqref{eq:src-nucleon-is-relations}, or the $pn$ pair dominance, this fixes $X = Y$. The factor $X$, which acts as a global scaling for the $C^A_p$, $C^A_n$ coefficients, is nevertheless still free. This means that for any given set of $C^A_p$, $C^A_n$, $f^{\mathrm{SRC}p}_i(x,Q)$, $f^{\mathrm{SRC}n}_i(x,Q)$ and $f^p_i(x,Q)$, $f^n_i(x,Q)$, it is possible to find an alternative SRC PDF parametrization $\tilde{f}^{\mathrm{SRC}p}_i(x,Q)$, $\tilde{f}^{\mathrm{SRC}n}_i(x,Q)$ that produces exactly the same full-nucleus PDFs $f^A_i(x,Q)$ with arbitrarily large or small coefficients $\tilde{C}^A_p$, $\tilde{C}^A_n$. It is thus \emph{not} possible to extract the $C^A_p$, $C^A_n$ coefficients individually from an nPDF analysis in a parametrization independent way. The overall normalization of the values obtained in Ref.~\cite{PhysRevLett.133.152502} is therefore necessarily specific to the SRC PDF parametrization given in the Supplemental Material of that Letter. We further stress this point by noting that if one would parametrize the SRC PDFs in a form of $f^{\mathrm{SRC}p}_i(x,Q) = f^p_i(x,Q) + \delta f^p_i(x,Q)$, with $\delta f^p_i(x,Q)$ an arbitrary function up to sum-rule requirements---bearing in mind that a fully general parametrization of the SRC PDFs could always be written in this form---the global normalizations of $C^A_p$ and $\delta f^p_i(x,Q)$ would be perfectly anticorrelated and thus unconstrainable in the fit.

Importantly, since the rescaling is global, applying to all nuclei simultaneously, it leaves the ratios $C^{A_1}_p/C^{A_2}_p$, $C^{A_1}_n/C^{A_2}_n$ of any two nuclei invariant. These ratios, or equivalently the $A$-dependence of the coefficients, can thus be extracted uniquely in a global analysis within the two-component assumption, even if the global normalisation remains unconstrainable. There is a corresponding finding in the nuclear-structure studies with ab-initio quantum Monte Carlo (QMC) calculations, where different nucleon-nucleon interactions yield different two-nucleon densities, known as the scheme (interaction-model) and scale (cut-off regulation) dependence of these calculations~\cite{Cruz-Torres:2019fum}. This dependence can be also understood in terms of the renormalization group evolution of the effective-field-theory resolution scale~\cite{PhysRevC.104.034311}. As a consequence, also the nucleus-dependent nuclear contact terms, which define the number of SRC pairs in the generalized contact formalism, can depend on the chosen interaction model. However, the ratios of different-nuclei contact terms show independence of the scheme and scale within the theory uncertainties~\cite{Cruz-Torres:2019fum}. The natural point of comparison between the SRC PDF abundance coefficients and the nuclear-theory contact-term predictions is therefore at the level of these $C^{A_1}_p/C^{A_2}_p$, $C^{A_1}_n/C^{A_2}_n$ ratios. Also experimentally, the cross sections for QE scattering are typically provided in terms of ratios to light nuclei, providing a common point of reference.

\section{Relation to nuclear modification factor}

The different nPDF parametrizations are often discussed in terms of the nuclear modification factors for the partons in bound protons (neutrons),
\begin{equation}
  R^{p(n)/A}_i(x,Q) = \frac{f^{p(n)/A}_i(x,Q)}{f^{p(n)}_i(x,Q)}.
\end{equation}
Similarly, one can define the modification factor for the partons in a SRC-pair nucleon,
\begin{equation}
  R^{\mathrm{SRC}p(n)}_i(x,Q) = \frac{f^{\mathrm{SRC}p(n)}_i(x,Q)}{f^{p(n)}_i(x,Q)}.
\end{equation}
From Eq.~\eqref{eq:bound-nucleon-from-src} it then follows that
\begin{equation}
  R^{p(n)/A}_i(x,Q) = 1 + C^A_{p(n)} \, [R^{\mathrm{SRC}p(n)}_i(x,Q) - 1]
  \label{eq:nucl_mod_fact_from_src}
\end{equation}
in the SRC-motivated parametrization. Given the universality of the SRC PDFs, we immediately see that the SRC decomposition requires
\begin{equation}
  \frac{R^{p(n)/A_1}_i(x,Q) - 1}{R^{p(n)/A_2}_i(x,Q) - 1} = \frac{C^{A_1}_{p(n)}}{C^{A_2}_{p(n)}},
  \label{eq:nucl_mod_fact_ratios_from_src}
\end{equation}
where the left-hand side depends on the parton flavor $i$, $x$ and $Q$, but the right-hand side does not. Thus some, but not all, nPDF parametrizations comply with the SRC-type decomposition. This is now a strong prediction of the two-component hypothesis of the SRC parametrization, and has an impact also on structure-function and cross-section ratios as shown below.

We cannot help but note the similarity of Eq.~\eqref{eq:nucl_mod_fact_from_src} with the EPPS21 parametrization~\cite{Eskola:2021nhw} where the $A$-dependence is given at the parametrization scale $Q_0$ for the strength of the nuclear modification at the EMC minimum $x_e$ and antishadowing maximum $x_a$ (and for sea quarks also at the limit $x \to 0$) as
\begin{equation}
  R^{p/A}_i(x_j,Q_0) = 1 + \left(\frac{A}{A_\mathrm{ref}}\right)^{\gamma_{\,i,j}} \, [R^{p/A_\mathrm{ref}}_i(x_j,Q_0) - 1]
  \label{eq:epps_mass_dependence}
\end{equation}
for $x_j \in {x_e,x_a}$ (and for $x_j = 0$ when $i = \bar{q}$). The SRC-motivated approach and EPPS21 can thus be seen as two examples of phenomenologically motivated nPDF parametrizations. There are a couple of important differences, though. First, importantly, the reference nucleus $A_\mathrm{ref}$ in the EPPS21 fit is completely arbitrary, and carries no physical meaning. It can be changed freely, $A_\mathrm{ref} \to \tilde{A}_\mathrm{ref}$, without changing the results for any physical observables, as Eq.~\eqref{eq:epps_mass_dependence} then simply relates $R^{p/\tilde{A}_\mathrm{ref}}_i(x_j,Q_0) = 1 + (\tilde{A}_\mathrm{ref}/A_\mathrm{ref})^{\gamma_{\,i,j}} \, [R^{p/A_\mathrm{ref}}_i(x_j,Q_0) - 1]$. This is similar to the rescaling invariance discussed above, but since this freedom relates to the choice of a quantity that is not given any special physical importance, $A_\mathrm{ref}$, which is not (and should not be) taken as a free parameter, there is no potential danger. Second, the parameters $\gamma_{\,i,j}$ are allowed to depend on the flavor $i$ and the point $x_j$ (though the lack of sufficient data required fixing the flavor dependence in some instances). This also means that the flavor mixing in the DGLAP evolution can make the $A$-dependence different at different scales. In this sense the EPPS21 parametrization is more general than the SRC-motivated approach. Third, the nuclear-mass dependence in EPPS21 follows a monotonous power-law behavior (except for the lightest nuclei, where additional parameters are applied for $^3\mathrm{He}$ and $^6\mathrm{Li}$), whereas the SRC-motivated approach is more general in this sense, since the abundance coefficients are allowed to vary independently from nucleus to nucleus.

\section{Reformulation of SRC-motivated parametrization}

As shown above, the parametrization in Eq.~\eqref{eq:bound-nucleon-from-src} is problematic due to the rescaling invariance, which prevents the extraction of the global normalization of the SRC abundance coefficients if a fully general form of the SRC PDFs is assumed. We propose here an alternative approach which circumvents this problem. Similarly to the EPPS21 parametrization discussed above, the two-component parametrization can be equivalently formulated in terms of a reference nucleus $A_\mathrm{ref}$ as
\begin{equation}
  \begin{split}
    f^{p/A}_i(x,Q) &= \left(1-\frac{C^A_p}{C^{A_\mathrm{ref}}_p}\right) \times f^p_i(x,Q) + \frac{C^A_p}{C^{A_\mathrm{ref}}_p} \times f^{p/A_\mathrm{ref}}_i(x,Q), \\
    f^{n/A}_i(x,Q) &= \left(1-\frac{C^A_n}{C^{A_\mathrm{ref}}_n}\right) \times f^n_i(x,Q) + \frac{C^A_n}{C^{A_\mathrm{ref}}_n} \times f^{n/A_\mathrm{ref}}_i(x,Q),
  \end{split}
  \label{eq:bound-nucleon-from-src-ref-nucleus}
\end{equation}
with Eq.~\eqref{eq:bound-nucleon-from-src-ref-nucleus} following directly from Eq.~\eqref{eq:bound-nucleon-from-src} by writing it in the case of $A$ and $A_\mathrm{ref}$ and solving the system of equations to get rid of the direct dependence on the SRC PDFs. Since only ratios of the SRC-abundance coefficients appear, this formulation is free from the rescaling-invariance problem. One can then parametrize the functions $f^{p/A_\mathrm{ref}}_i(x,Q)$ and $f^{n/A_\mathrm{ref}}_i(x,Q)$ freely, but respecting the sum rules, and fit to data together with the $C^A_p/C^{A_\mathrm{ref}}_p$, $C^A_n/C^{A_\mathrm{ref}}_n$ ratios. The treatment of IS, assuming Eq.~\eqref{eq:src-nucleon-is-relations} holds, is simplified if one chooses $A_\mathrm{ref}$ to be an isoscalar nucleus, as then one can set $C^{A_\mathrm{ref}}_n = C^{A_\mathrm{ref}}_p$, $f^{n/A_\mathrm{ref}}_u(x,Q) = f^{p/A_\mathrm{ref}}_d(x,Q)$, $f^{n/A_\mathrm{ref}}_d(x,Q) = f^{p/A_\mathrm{ref}}_u(x,Q)$, etc., but it is also possible to use a non-isoscalar nucleus as a reference by setting
\begin{equation}
  \begin{split}
    f^{n/A_\mathrm{ref}}_u(x,Q) &= f^{p/A_\mathrm{ref}}_d(x,Q) - \left(1 - \frac{C^{A_\mathrm{ref}}_n}{C^{A_\mathrm{ref}}_p}\right) \, [f^{p/A_\mathrm{ref}}_d(x,Q) - f^p_d(x,Q)], \\
    f^{n/A_\mathrm{ref}}_d(x,Q) &= f^{p/A_\mathrm{ref}}_u(x,Q) - \left(1 - \frac{C^{A_\mathrm{ref}}_n}{C^{A_\mathrm{ref}}_p}\right) \, [f^{p/A_\mathrm{ref}}_u(x,Q) - f^p_u(x,Q)],
  \end{split}
\end{equation}
with the ratio $C^{A_\mathrm{ref}}_n/C^{A_\mathrm{ref}}_p$ as a free parameter.

The parametrization in Eq.~\eqref{eq:bound-nucleon-from-src-ref-nucleus} shares all the same properties as the original one in Eq.~\eqref{eq:bound-nucleon-from-src}, for example Eq.~\eqref{eq:nucl_mod_fact_ratios_from_src} follows directly, but without any reference to the SRC PDFs. These can be recovered, under the SRC interpretation, through
\begin{equation}
  f^{\mathrm{SRC}p(n)}_i(x,Q) = \frac{1}{C^A_{p(n)}}[f^{p(n)/A}_i(x,Q) - (1-C^A_{p(n)})f^{p(n)}_i(x,Q)],
\end{equation}
where it should not matter which $A$ one uses. However, since only the ratios $C^A_p/C^{A_\mathrm{ref}}_p$, $C^A_n/C^{A_\mathrm{ref}}_n$ are extracted in the fit, one must fix the coefficients $C^A_{p(n)}$ for a single nucleus in order to get the $f^{\mathrm{SRC}p(n)}_i(x,Q)$. This emphasises that model-dependent information is needed in order to extract the SRC PDFs from an nPDF fit.

\section{Scaling relations for structure-function ratios}

From the SRC-motivated parametrization, it follows that \emph{any} structure function (or cross section) linear in nPDFs, $F^A_i(x,Q) = \sum_j C_{ij}(Q) \otimes f^A_j(Q)$, keeping for convenience the per-nucleon normalization, can be written as
\begin{equation}
  F^A_i(x,Q) = \frac{Z}{A} [(1-C^A_p) \times F^p_i(x,Q) + C^A_p \times F^{\mathrm{SRC}p}_i(x,Q)] + \frac{N}{A} [(1-C^A_n) \times F^n_i(x,Q) + C^A_n \times F^{\mathrm{SRC}n}_i(x,Q)].
\end{equation}
In particular, for the isoscalar nuclei this implies, using Eq.~\eqref{eq:isoscalar-coeff-is-relation} and neglecting the (percent-level) deuteron corrections, that
\begin{equation}
  \frac{F^{Z=N}_i(x,Q)}{F^\mathrm{D}_i(x,Q)} = 1 + C^{(Z=N)}_p \left[ \frac{F^{\mathrm{SRC}p}_i(x,Q) + F^{\mathrm{SRC}n}_i(x,Q)}{F^p_i(x,Q) + F^n_i(x,Q)} - 1 \right].
  \label{eq:sf_ratio_src}
\end{equation}
This matches exactly with the $F^A_2$ structure function scaling which was demonstrated to hold for nuclear-DIS data already 40 years ago in Refs.~\cite{PhysRevLett.52.2344,FRANKFURT1988235}. In Ref.~\cite{PhysRevLett.52.2344} this scaling is expressed for isoscalarized data (modifying the notation slightly) as
\begin{equation}
  \frac{F^{Z=N}_2(x,Q)}{F^\mathrm{D}_2(x,Q)} = 1 + P_\mathrm{ex}(A) [R_\mathrm{ex}(x) - 1],
  \label{eq:sf_ratio_exotic}
\end{equation}
where $P_\mathrm{ex}(A)$ is the probability of finding an exotic component in the nucleus $A$ and $R_\mathrm{ex}(x)$ is the ratio of the $F^\mathrm{ex}_2$ structure function for the exotic component with respect to that for an isoscalarized free nucleon. The expressions on the right-hand sides of Eqs.~\eqref{eq:sf_ratio_src} and~\eqref{eq:sf_ratio_exotic} are still subject to the rescaling invariance since the SRC and exotic component probabilities can be scaled together with the respective structure functions. Notably, when extracting the $A$-dependence of $P_\mathrm{ex}(A)$ from experimental data in Ref.~\cite{PhysRevLett.52.2344}, the normalization was fixed \emph{by hand}, setting $P_\mathrm{ex}(^4\mathrm{He}) = 0.12$ based on a multi-quark cluster model of the exotic component. However, it was stated that the scaling property of Eq.~\eqref{eq:sf_ratio_exotic} itself should not depend on the normalization of $P_\mathrm{ex}(A)$.

Ref.~\cite{PhysRevLett.52.2344} further notes that the $A$-dependence of $P_\mathrm{ex}(A)$ can be derived from
\begin{equation}
  \frac{F^{Z_1=N_1}_2(x,Q)/F^\mathrm{D}_2(x,Q) - 1}{F^{Z_2=N_2}_2(x,Q)/F^\mathrm{D}_2(x,Q) - 1} = \frac{P_\mathrm{ex}(A_1)}{P_\mathrm{ex}(A_2)}.
\end{equation}
Intriguingly, it was noted in Ref.~\cite{PhysRevLett.52.2344} that the best agreement with the extracted $A$-dependence was reached within the considered model if six-quark clusters were assumed as the additional component, clearly resembling the local-density picture of SRCs~\cite{PhysRevC.86.065204,PhysRevLett.123.042501}. For the SRC abundances we similarly find
\begin{equation}
  \frac{F^{Z_1=N_1}_i(x,Q)/F^\mathrm{D}_i(x,Q) - 1}{F^{Z_2=N_2}_i(x,Q)/F^\mathrm{D}_i(x,Q) - 1} = \frac{C^{(Z_1=N_1)}_p}{C^{(Z_2=N_2)}_p} = \frac{C^{(Z_1=N_1)}_n}{C^{(Z_2=N_2)}_n}.
  \label{eq:src-scaling-dis-sf}
\end{equation}
As noted above, the coefficient ratios in Eq.~\eqref{eq:src-scaling-dis-sf} are invariant under the rescaling in Eq.~\eqref{eq:src-rescaling-trasformation}, as they should since the left-hand side depends only on observable quantities. It is thus possible to extract the $A$-dependence of the abundance coefficient directly from data independently of the SRC PDF parametrization, but as we have discussed above, not their normalization.

The scaling discussed above is the same as introduced in Ref.~\cite{CLAS:2019vsb} for the $F_2$ structure function within the $pn$ pair dominance assumption, but in lack of a direct relation between the $C^A_p$ and $C^A_n$ coefficients for non-isoscalar nuclei, we had to stick with isoscalar nuclei in the expressions. Allowing for a non-zero SRC component in the deuteron, Eq.~\eqref{eq:sf_ratio_src} generalises to
\begin{equation}
  \frac{F^{Z=N}_i(x,Q)}{F^\mathrm{D}_i(x,Q)} = 1 + \left(\frac{C^{(Z=N)}_p}{C^\mathrm{D}_p} - 1\right) \, f^\mathrm{univ}_i(x,Q),
  \label{eq:sf_ratio_src_w_deut}
\end{equation}
where
\begin{equation}
  f^\mathrm{univ}_i(x,Q) =  C^\mathrm{D}_p \, \frac{F^{\mathrm{SRC}p}_i(x,Q) - F^p_i(x,Q) + F^{\mathrm{SRC}n}_i(x,Q) - F^n_i(x,Q)}{F^\mathrm{D}_i(x,Q)}
\end{equation}
is, for $i = 2$, the universal function introduced in Refs.~\cite{CLAS:2019vsb,PhysRevLett.124.092002}. By introducing the $pn$ pair dominance, Eq.~\eqref{eq:sf_ratio_src_w_deut} further generalises to all nuclei with
\begin{equation}
  \frac{F^A_i(x,Q)}{F^\mathrm{D}_i(x,Q)} = \frac{2N}{A} + \frac{2(Z - N)}{A} \, \frac{F^p_i(x,Q)}{F^\mathrm{D}_i(x,Q)} + \left(\frac{n^A_{pn}}{n^\mathrm{D}_{pn}} - \frac{2N}{A}\right)\, f^\mathrm{univ}_i(x,Q),
  \label{eq:sf_ratio_src_pnd}
\end{equation}
which is the form presented in Refs.~\cite{CLAS:2019vsb,PhysRevLett.124.092002}. Notably, $f^\mathrm{univ}_i(x,Q)$ is invariant under the rescaling transformation, and thus its extraction in Ref~\cite{PhysRevLett.124.092002} for the $i = 2$ case could be uniquely done. This follows from fixing the deuteron as the reference nucleus, after which only ratios of the abundance coefficients to those of the deuteron appear in the formulae together with the universal function.

Importantly, due to the flavor independence of the abundance coefficients, Eqs.~\eqref{eq:src-scaling-dis-sf}, \eqref{eq:sf_ratio_src_w_deut}, and~\eqref{eq:sf_ratio_src_pnd} hold for all structure functions, and not only for $F_2$ discussed in the earlier works~\cite{CLAS:2019vsb,PhysRevLett.124.092002,PhysRevLett.52.2344,FRANKFURT1988235}. Refs.~\cite{PhysRevD.104.033002,PhysRevD.108.053005} appear to be theory-level demonstrations of this scaling in the case of neutrino-nucleus DIS, which was therefore suggested as a test of SRC universality, but the limited amount of data for this process makes it difficult to test in practice. The limits of this scaling could be put to a stringent test in an extended kinematic range with the EIC, where it becomes possible to test it also for the longitudinal structure function and charm-production DIS cross section~\cite{PhysRevD.96.114005}. In particular, the two-component hypothesis, if taken literally, predicts that Eq.~\eqref{eq:src-scaling-dis-sf} yields the same ratio for the $F^A_2$, $F^A_\mathrm{L}$ structure functions and for the charm-production DIS cross section, even at different kinematics (cf.~Ref.~\cite{Xu:2019wso}).

\section{Scaling relations for hadron-nucleus cross sections}

Since the structure-function scaling extends to any cross section with linear dependence on the nPDFs, it applies also for hadron-nucleus ($h+A$) collisions. The cross section for producing a (semi-)inclusive final state $k+X$, $\sigma_{h+A}^{k+X} = \sum_{i,j} f^h_i \otimes \hat{\sigma}_{i+j}^{k+X} \otimes f^A_j$, can thus be written as
\begin{equation}
  \sigma_{h+A}^{k+X} = \frac{Z}{A} \left[(1-C^A_p) \times \sigma_{h+p}^{k+X} + C^A_p \times \sigma_{h+\mathrm{SRC}p}^{k+X}\right] + \frac{N}{A} \left[(1-C^A_n) \times \sigma_{h+n}^{k+X} + C^A_n \times \sigma_{h+\mathrm{SRC}n}^{k+X}\right].
\end{equation}
One can see that this leads to a similar scaling relation
\begin{equation}
  \frac{\sigma_{h+A_1}^{k+X}/\sigma_{h+\mathrm{D}}^{k+X} - 1}{\sigma_{h+A_2}^{k+X}/\sigma_{h+\mathrm{D}}^{k+X} - 1} = \frac{C^{(Z_1=N_1)}_p}{C^{(Z_2=N_2)}_p} = \frac{C^{(Z_1=N_1)}_n}{C^{(Z_2=N_2)}_n}
  \label{eq:src-scaling-hA-xsec}
\end{equation}
for isoscalar nuclei as in the case of DIS structure functions. This expression can of course be extended to include non-isoscalarity and deuteron corrections, paticularly in the case of $pn$ pair dominance. Very recently, Ref.~\cite{Huang:2025kmd} suggested this scaling for the pion-nucleus Drell-Yan (DY) processes, but we note that it extends to \emph{any} collinearly factorizable hadron-nucleus cross section linear in nPDFs. Eqs.~\eqref{eq:src-scaling-dis-sf} and~\eqref{eq:src-scaling-hA-xsec} relate directly, as a consequence of the flavor-independent nuclear-mass dependence, the scaling of DIS and hadronic cross sections to each other. This uniform scaling appears to be consistent with the limited amount of fixed-target DIS and DY data available for the global fits, but could be tested in a wider kinematical region by comparing future DIS data from the EIC to DY measurements at the LHC, but only if a light-ion program at the latter is realised with a variety of nuclei and sufficient statistics.

For processes with a relatively weak isospin dependence, such as jet production~\cite{Eskola:2013aya}, one finds an approximate scaling relation that holds also for non-isoscalar nuclei
\begin{equation}
  \frac{\sigma_{p+A}^{k+X}}{\sigma_{p+p}^{k+X}} \approx 1 + \left(\frac{Z}{A}C^A_p + \frac{N}{A}C^A_n\right) \left[\frac{\sigma_{p+\mathrm{SRC}p}^{k+X}}{\sigma_{p+p}^{k+X}} - 1\right].
\end{equation}
We thus see that the two-component approach predicts e.g.\ that
\begin{equation}
  \frac{\sigma_{p+\mathrm{Pb}}^{\mathrm{dijet}+X}/\sigma_{p+p}^{\mathrm{dijet}+X} - 1}{\sigma_{p+\mathrm{O}}^{\mathrm{dijet}+X}/\sigma_{p+p}^{\mathrm{dijet}+X} - 1} \approx \frac{\frac{82}{208}C^\mathrm{Pb}_p + \frac{126}{208}C^\mathrm{Pb}_n}{\frac{8}{16}C^\mathrm{O}_p + \frac{8}{16}C^\mathrm{O}_n},
  \label{eq:lead-oxygen-dijet-ratio}
\end{equation}
where the right-hand side is a constant that is independent of the dijet kinematics, and the non-isoscalarity corrections to this ratio should be very small. This prediction could be tested with the single-differential dijet-production measurement that should be possible with the short $p+\mathrm{O}$ run at LHC~\cite{PhysRevD.105.L031504}, if the collision energies for $p+\mathrm{O}$, $p+\mathrm{Pb}$ and a $p+p$ reference are matched by interpolation. Within the $pn$ pair dominance assumption the right-hand side of Eq.~\eqref{eq:lead-oxygen-dijet-ratio} reduces to $(n^\mathrm{Pb}_{pn}/208) / (n^\mathrm{O}_{pn}/16)$.

\section{Alternative interpretation of extracted abundance coefficients}

As we have discussed above, only the $A$-dependence of the abundance coefficients can be meaningfully constrained in an nPDF fit. In addition, the two-component parametrization is quite general and open for different interpretations. The natural question to follow is then whether the available experimental data give evidence of an $A$-dependence that would be unique to the SRC interpretation, or if some other model could explain the extracted coefficients. To demonstrate this point, we entertain the possibility that the modified nucleons in the two-component approach do not necessarily come in SRC pairs, but rather reside more likely in the center of a nucleus than on its surface, irrespectively of the modification mechanism. The fraction of modified nucleons in the nucleus should therefore have a constant volume term and a surface correction proportional to $A^{-1/3}$.

In more detail, following Ref.~\cite{PhysRevLett.52.2344}, we define the central region as the volume $V_\mathrm{C} = \frac{4\pi}{3} R_\mathrm{C}^3$ confined within a radius $R_\mathrm{C} = R - T/2$, where $R \approx 1.12\ \mathrm{fm} \times A^{1/3}$ is the radius of the nucleus and $T$ is the approximately $A$-independent surface thickness (proportional to the diffuseness parameter of the Fermi distribution). Within this central region, the nucleon density is approximately constant $\rho_0 \approx 0.17\ \mathrm{fm}^{-3}$, and the proton and neutron densities can be assumed to follow their relative numbers in the full nucleus, $\rho_{p,\mathrm{C}} / \rho_0 = Z / A$ and $\rho_{n,\mathrm{C}} / \rho_0 = N / A$ (cf.\ Ref.~\cite{PhysRevC.76.014311}). The total number of protons within the central volume is therefore
\begin{equation}
  n_{p,\mathrm{C}} = \rho_{p,\mathrm{C}} \, V_\mathrm{C} = Z \times \frac{\rho_0 \, V_\mathrm{C}}{A},
\end{equation}
the remaining number being contained on the surface
\begin{equation}
  n_{p,\mathrm{S}} = Z - n_{p,\mathrm{C}} = Z \times \left( 1 - \frac{\rho_0 \, V_\mathrm{C}}{A} \right),
\end{equation}
and the fraction of nucleons in the central volume can be approximated as
\begin{equation}
  \frac{\rho_0 \, V_\mathrm{C}}{A} = \frac{4\pi}{3} \, \frac{\rho_0 \, R_\mathrm{C}^3}{A} \approx 1 - \frac{3T}{2 \times 1.12\ \mathrm{fm}} A^{-1/3}.
\end{equation}
We now employ the assumption that the probability for a proton being modified is different in the central region ($P^\mathrm{mod}_{p,\mathrm{C}}$) than on the surface ($P^\mathrm{mod}_{p,\mathrm{S}}$). The total number of modified protons is then
\begin{equation}
  n^\mathrm{mod}_p = P^\mathrm{mod}_{p,\mathrm{C}} \, n_{p,\mathrm{C}} + P^\mathrm{mod}_{p,\mathrm{S}} \, n_{p,\mathrm{S}} = Z \times \left[ P^\mathrm{mod}_{p,\mathrm{C}} \, \frac{\rho_0 \, V_\mathrm{C}}{A} + P^\mathrm{mod}_{p,\mathrm{S}} \, \left( 1 - \frac{\rho_0 \, V_\mathrm{C}}{A} \right) \right] \approx Z \times (C_p^\mathrm{V} + C_p^\mathrm{S}A^{-1/3}),
\end{equation}
where the volume and surface-correction coefficients are
\begin{equation}
  \begin{split}
    C_p^\mathrm{V} &= P^\mathrm{mod}_{p,\mathrm{C}}, \\
    C_p^\mathrm{S} &= \frac{3T}{2 \times 1.12\ \mathrm{fm}} \, (P^\mathrm{mod}_{p,\mathrm{S}} - P^\mathrm{mod}_{p,\mathrm{C}}).
  \end{split}
\end{equation}
Following the same reasoning for the number of modified neutrons, and using the per-nucleon definitions of the abundance coefficients $C_p^A = n^\mathrm{mod}_p / Z$, $C_n^A = n^\mathrm{mod}_n / N$, we arrive with
\begin{equation}
  \begin{split}
    C_p^A &= C_p^\mathrm{V} + C_p^\mathrm{S} \, A^{-1/3}, \\
    C_n^A &= C_n^\mathrm{V} + C_n^\mathrm{S} \, A^{-1/3}.
  \end{split}
\end{equation}

\begin{figure}
  \includegraphics[width=\textwidth]{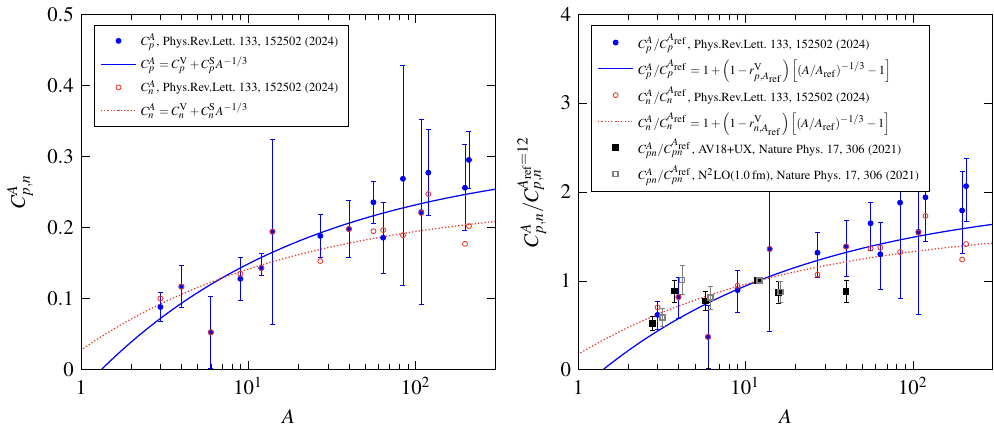}
  \caption{A demonstration of the volume + surface term scaling of the abundances extracted in Ref.~\cite{PhysRevLett.133.152502}. Left: Two-parameter fits to the $C_p^A$, $C_n^A$ coefficients. Right: One-parameter fits for the ratios $C_p^A / C_p^{A_\mathrm{ref}}$, $C_n^A / C_n^{A_\mathrm{ref}}$, and a comparison to spin-1 $pn$ contact-term ratios obtained from coordinate-space extraction of two different ab-initio QMC calculations in Ref.~\cite{Cruz-Torres:2019fum}. The possible correlations across different nuclei are neglected in constructing the uncertainties in the ratios.}
  \label{fig:CA-fit}
\end{figure}

A two-parameter fit to the $C_p^A$ values presented in Ref.~\cite{PhysRevLett.133.152502}, neglecting their possible correlations, yields the values $C_p^\mathrm{V} = 0.304$ and $C_p^\mathrm{S} = -0.334$ with a $\chi^2 = 7.27$ figure of merit for 13 degrees of freedom. The fit results are presented in FIG.~\ref{fig:CA-fit} (left-hand-side plot), where we see that this form indeed nicely describes the extracted $C_p^A$ coefficients throughout the $A$ range within the quoted uncertainties. For the $C_n^A$ coefficients we cannot do the fit as easily since Ref.~\cite{PhysRevLett.133.152502} does not provide their errors directly. It is however mentioned that the uncertainties are similar in size, so as a crude approximation, we can assume that the $C_n^A$ errors are exactly the same as the $C_p^A$ errors, nucleus by nucleus. This yields $C_n^\mathrm{V} = 0.240$ and $C_n^\mathrm{S} = -0.213$ with a $\chi^2 = 4.07$, again nicely describing the $A$-dependence.

One might be inclined to infer the central-region and surface modification probabilities from the fitted values, but as we have stressed above, the normalization of the $C^A_p$, $C^A_n$ coefficients is subject to the rescaling invariance, and thus also that of the fitted volume and surface coefficients can be changed arbitrarily. In fact, we should only study the ratios of the abundance coefficients, e.g.\ with respect to a single reference nucleus $A_\mathrm{ref}$. These may be written as
\begin{equation}
  \begin{split}
    \frac{C_p^A}{C_p^{A_\mathrm{ref}}} &= 1 + \left( 1 - r^{\mathrm{V}}_{p,A_\mathrm{ref}} \right) \left[ \left(\frac{A}{A_\mathrm{ref}}\right)^{-1/3} - 1 \right], \\
    \frac{C_n^A}{C_n^{A_\mathrm{ref}}} &= 1 + \left( 1 - r^{\mathrm{V}}_{n,A_\mathrm{ref}} \right) \left[ \left(\frac{A}{A_\mathrm{ref}}\right)^{-1/3} - 1 \right],
  \end{split}
  \label{eq:one-parameter-fit-form}
\end{equation}
where only one free parameter then remains for both nucleon types, with following relations for the protons (neutrons),
\begin{equation}
  r^{\mathrm{V}}_{p(n),A_\mathrm{ref}} = \frac{C_{p(n)}^\mathrm{V}}{C_{p(n)}^\mathrm{V} + C_{p(n)}^\mathrm{S} \, A_\mathrm{ref}^{-1/3}} = \frac{1}{1 + \frac{3T}{2 \times 1.12\ \mathrm{fm}} \, (P^\mathrm{mod}_{p(n),\mathrm{S}} / P^\mathrm{mod}_{p(n),\mathrm{C}} - 1) \, A_\mathrm{ref}^{-1/3}}.
\end{equation}
Fitting these parameters using the functional form in Eq.~\eqref{eq:one-parameter-fit-form} and the coefficient ratios obtained from Ref.~\cite{PhysRevLett.133.152502} with $A_\mathrm{ref} = 12$, we obtain $r^{\mathrm{V}}_{p,A_\mathrm{ref}} = 1.96$ and $r^{\mathrm{V}}_{n,A_\mathrm{ref}} = 1.64$, with $\chi^2$ values identical to the two-parameter-fit case. The results are illustrated in FIG.~\ref{fig:CA-fit} (right-hand-side plot). It should be emphasised that only the ratios of surface and central-region modification probabilities can be inferred from this fit (already a model-dependent extraction), and fixing their absolute size would require further input, just as for the abundance coefficients in the SRC interpretation.

It appears that the abundances extracted in Ref.~\cite{PhysRevLett.133.152502} simply demonstrate the ``unreasonably good'' agreement of the PDF nuclear modifications with the volume plus surface term scaling even down to very low nuclear masses~\cite{PhysRevC.86.065204}. Importantly, thanks to the work in Ref.~\cite{PhysRevLett.133.152502}, this scaling is now derived from a global analysis of nPDFs, within the two-component approach, in contrast to the earlier analyses restricted to $F_2$ structure functions. In showing this scaling, we did not need to assume the modified component to be of SRC origin. We thus note that while it is possible to interpret the extracted $C^A_p$, $C^A_n$ coefficients as SRC-nucleon abundances, this is not the only viable interpretation. Intriguingly, though, the modified-proton and neutron abundance coefficients appear to have a different slope with respect to $A$, which in Ref.~\cite{PhysRevLett.133.152502} was interpreted as a consequence of a $pn$ pair dominated SRC component. However, the same apparent dependence could also be an effect of an isovector mean-field modification~\cite{PhysRevLett.102.252301}. A more detailed comparison of predictions from these two models would therefore be warranted. For completeness, we show in FIG.~\ref{fig:CA-fit} (right-hand-side plot) also a comparison to the generalized contact formalism results from QMC calculations~\cite{Cruz-Torres:2019fum} in a similar fashion to Ref.~\cite{PhysRevLett.133.152502}, but now in terms of the ratios with respect to a reference nucleus. These are in a reasonable agreement with the abundance-coefficient ratios extracted from Ref.~\cite{PhysRevLett.133.152502}, but a definitive comparison would require extending the QMC calculations to higher nuclear masses and reducing the uncertainties in the nPDF global-analysis extraction.

\section{Conclusion}

Concluding, the nuclear modifications in PDFs may, or may not, be caused by nucleon binding in SRC pairs. As discussed in Refs.~\cite{PhysRevLett.133.152502,RevModPhys.89.045002}, it is a plausible explanation particularly for the large-$x$ EMC effect. Establishing this as \emph{the} correct interpretation is rather difficult since the inclusive hard processes where PDFs are defined probe only the partonic structure of nuclei without a direct sensitivity to the lower scale nucleonic degrees of freedom, and conversely the QE processes where the inter-nucleon dynamics can be accessed are principally indifferent of the sub-nucleonic structure. To connect these phenomena, Ref.~\cite{PhysRevLett.133.152502} uses the SRC-motivated nPDF parametrization given in Eq.~\eqref{eq:bound-nucleon-from-src}. We credit Ref.~\cite{PhysRevLett.133.152502} for taking this kind of a two-component model to the PDF global-analysis level, facilitating data comparisons beyond DIS observables as performed in their fit. We nevertheless stress that the same values of hard-process cross sections can be achieved either with a small number of strongly modified SRC nucleon pairs or with a large number of weakly modified pairs, and due to the rescaling invariance shown above, it is not possible to extract the $C^A_p$, $C^A_n$ coefficients uniquely from an nPDF fit. Their ratios $C^{A_1}_p/C^{A_2}_p$, $C^{A_1}_n/C^{A_2}_n$ are invariant under this rescaling transformation, though, which then allows for extracting these ratios from a fit and testing against the QE scattering data and nuclear theory predictions. We thus argue that any evidence for the connection between the SRCs and nuclear modifications of PDFs should be derived only from the $A$-dependence of the ratios of these coefficients, not from their absolute normalization.

We have further shown that the SRC-motivated nPDF parametrization naturally yields (as anticipated~\cite{FRANKFURT1988235}) the $F^A_2$ structure function scaling for isoscalar nuclei found in Ref.~\cite{PhysRevLett.52.2344}, but---due to the flavor-independent $A$-dependence---holds also for any other structure function or cross section linear in nPDFs. The SRC two-component hypothesis thus has predictive power and can be tested with a variety of different nuclei and DIS cross sections at the EIC. This nuclear scaling could be tested also at the LHC, but only if a sufficiently broad light-ion program is realised. While these scaling relations necessarily follow from assuming the SRCs as the dominant cause for the nuclear modifications of PDFs, the opposite might not be true, as the two-component decomposition of the nuclear effects is still quite general and open for different interpretations. The relevant question then is whether SRCs predict a nuclear mass dependence of the nPDFs distinct from any other explanation. With the global data and the extracted $C^A_p$, $C^A_n$ coefficients being currently equally well described in terms of the SRC framework, or, as shown above, in any other model with approximately volume plus surface correction type scaling, clearly more work is needed to pin down the mechanism behind the nuclear modifications of PDFs. A possible way forward is to test the scaling behaviour predicted by the two-component assumption for a range of observables and constraining the $C^{A_1}_p/C^{A_2}_p$, $C^{A_1}_n/C^{A_2}_n$ ratios, e.g.\ through the reformulated parametrization given above for circumventing the rescaling problem, to a limit precise enough to differentiate between models.

\vspace{-0.5cm}
\begin{acknowledgments}
  The author wishes to thank Tomas Je\v{z}o, Aleksander Kusina, Fredrick Olness and Peter Risse for discussions concerning the SRC PDF approach and the parametrization used in Ref.~\cite{PhysRevLett.133.152502}, Or Hen for pointing out the connection with the scheme and scale dependence in the ab-initio QMC calculations, and Hannu Paukkunen for further discussions and comments. The work of P.P.\ has been funded through the Research Council of Finland (projects No.\ 330448 and No.\ 331545), as a part of the Center of Excellence in Quark Matter of the Research Council of Finland (project No.\ 364194), and as a part of the  European Research Council project No.\ ERC-2018-ADG-835105 YoctoLHC.
\end{acknowledgments}

\vspace{-0.2cm}
\bibliography{comment-on-src-pdfs}

@book{Collins:2011zzd,
    author = "Collins, John",
    title = "{Foundations of Perturbative QCD}",
    doi = "10.1017/9781009401845",
    isbn = "978-1-009-40184-5, 978-1-009-40183-8, 978-1-009-40182-1",
    publisher = "Cambridge University Press",
    volume = "32",
    year = "2011"
}

@article{Dokshitzer:1977sg,
      author         = "Dokshitzer, Yuri L.",
      title          = "{Calculation of the Structure Functions for Deep Inelastic Scattering and e+ e- Annihilation by Perturbation Theory in Quantum Chromodynamics.}",
      journal        = "Sov. Phys. JETP",
      volume         = "46",
      year           = "1977",
      pages          = "641",
      note           = "[Zh. Eksp. Teor. Fiz. 73 (1977) 1216]"
}

@article{Gribov:1972ri,
      author         = "Gribov, V. N. and Lipatov, L. N.",
      title          = "{Deep inelastic e p scattering in perturbation theory}",
      journal        = "Sov. J. Nucl. Phys.",
      volume         = "15",
      year           = "1972",
      pages          = "438",
      note           = "[Yad. Fiz. 15 (1972) 781]"
}

@article{Lipatov:1974qm,
      author         = "Lipatov, L. N.",
      title          = "{The parton model and perturbation theory}",
      journal        = "Sov. J. Nucl. Phys.",
      volume         = "20",
      year           = "1975",
      pages          = "94",
      note           = "[Yad. Fiz. 20 (1974) 181]"
}

@article{Altarelli:1977zs,
      author         = "Altarelli, Guido and Parisi, G.",
      title          = "{Asymptotic Freedom in Parton Language}",
      journal        = "Nucl. Phys.",
      volume         = "B126",
      year           = "1977",
      pages          = "298"
}

@article{Arneodo:1992wf,
    author = "Arneodo, Michele",
    title = "{Nuclear effects in structure functions}",
    reportNumber = "CERN-PPE-92-113",
    doi = "10.1016/0370-1573(94)90048-5",
    journal = "Phys. Rept.",
    volume = "240",
    pages = "301--393",
    year = "1994"
}

@article{annurev-nucl-102122-022747,
  author = "Klasen, Michael and Paukkunen, Hannu",
  title = "{Nuclear Parton Distribution Functions After the First Decade of LHC Data}",
  journal= "Annual Review of Nuclear and Particle Science",
  year = "2024",
  volume = "74",
  number = "Volume 74, 2024",
  pages = "49-87",
  doi = "https://doi.org/10.1146/annurev-nucl-102122-022747",
  url = "https://www.annualreviews.org/content/journals/10.1146/annurev-nucl-102122-022747",
  publisher = "Annual Reviews",
  issn = "1545-4134",
  type = "Journal Article"
}

@article{PhysRevLett.133.152502,
  title = {{Modification of Quark-Gluon Distributions in Nuclei by Correlated Nucleon Pairs}},
  author = {Denniston, A. W. and Je\ifmmode \check{z}\else \v{z}\fi{}o, T. and Kusina, A. and Derakhshanian, N. and Duwent\"aster, P. and Hen, O. and Keppel, C. and Klasen, M. and Kova\ifmmode \check{r}\else \v{r}\fi{}\'{\i}k, K. and Morf\'{\i}n, J. G. and Muzakka, K. F. and Olness, F. I. and Piasetzky, E. and Risse, P. and Ruiz, R. and Schienbein, I. and Yu., J. Y.},
  journal = {Phys. Rev. Lett.},
  volume = {133},
  issue = {15},
  pages = {152502},
  numpages = {7},
  year = {2024},
  month = {Oct},
  publisher = {American Physical Society},
  doi = {10.1103/PhysRevLett.133.152502},
  url = {https://link.aps.org/doi/10.1103/PhysRevLett.133.152502}
}

@article{RevModPhys.89.045002,
  title = {Nucleon-nucleon correlations, short-lived excitations, and the quarks within},
  author = {Hen, Or and Miller, Gerald A. and Piasetzky, Eli and Weinstein, Lawrence B.},
  journal = {Rev. Mod. Phys.},
  volume = {89},
  issue = {4},
  pages = {045002},
  numpages = {49},
  year = {2017},
  month = {Nov},
  publisher = {American Physical Society},
  doi = {10.1103/RevModPhys.89.045002},
  url = {https://link.aps.org/doi/10.1103/RevModPhys.89.045002}
}

@article{PhysRevC.104.034311,
  title = {Short-range correlation physics at low renormalization group resolution},
  author = {Tropiano, A. J. and Bogner, S. K. and Furnstahl, R. J.},
  journal = {Phys. Rev. C},
  volume = {104},
  issue = {3},
  pages = {034311},
  numpages = {16},
  year = {2021},
  month = {Sep},
  publisher = {American Physical Society},
  doi = {10.1103/PhysRevC.104.034311},
  url = {https://link.aps.org/doi/10.1103/PhysRevC.104.034311}
}

@article{Cruz-Torres:2019fum,
    author = "Cruz-Torres, R. and Lonardoni, D. and Weiss, R. and Barnea, N. and Higinbotham, D. W. and Piasetzky, E. and Schmidt, A. and Weinstein, L. B. and Wiringa, R. B. and Hen, O.",
    title = "{Many-body factorization and position{\textendash}momentum equivalence of nuclear short-range correlations}",
    reportNumber = "LA-UR-19-25832",
    doi = "10.1038/s41567-020-01053-7",
    journal = "Nature Phys.",
    volume = "17",
    number = "3",
    pages = "306--310",
    year = "2021"
}

@article{PhysRevC.86.065204,
  title = {Detailed study of the nuclear dependence of the EMC effect and short-range correlations},
  author = {Arrington, J. and Daniel, A. and Day, D. B. and Fomin, N. and Gaskell, D. and Solvignon, P.},
  journal = {Phys. Rev. C},
  volume = {86},
  issue = {6},
  pages = {065204},
  numpages = {13},
  year = {2012},
  month = {Dec},
  publisher = {American Physical Society},
  doi = {10.1103/PhysRevC.86.065204},
  url = {https://link.aps.org/doi/10.1103/PhysRevC.86.065204}
}

@article{PhysRevLett.123.042501,
  title = {Searching for Flavor Dependence in Nuclear Quark Behavior},
  author = {Arrington, J. and Fomin, N.},
  journal = {Phys. Rev. Lett.},
  volume = {123},
  issue = {4},
  pages = {042501},
  numpages = {6},
  year = {2019},
  month = {Jul},
  publisher = {American Physical Society},
  doi = {10.1103/PhysRevLett.123.042501},
  url = {https://link.aps.org/doi/10.1103/PhysRevLett.123.042501}
}

@article{RevModPhys.82.2009,
  title = {Charge symmetry at the partonic level},
  author = {Londergan, J. T. and Peng, J. C. and Thomas, A. W.},
  journal = {Rev. Mod. Phys.},
  volume = {82},
  issue = {3},
  pages = {2009--2052},
  numpages = {0},
  year = {2010},
  month = {Jul},
  publisher = {American Physical Society},
  doi = {10.1103/RevModPhys.82.2009},
  url = {https://link.aps.org/doi/10.1103/RevModPhys.82.2009}
}

@article{PhysRevLett.102.252301,
  title = {{Isovector EMC Effect and the NuTeV Anomaly}},
  author = {Clo\"et, I. C. and Bentz, W. and Thomas, A. W.},
  journal = {Phys. Rev. Lett.},
  volume = {102},
  issue = {25},
  pages = {252301},
  numpages = {4},
  year = {2009},
  month = {Jun},
  publisher = {American Physical Society},
  doi = {10.1103/PhysRevLett.102.252301},
  url = {https://link.aps.org/doi/10.1103/PhysRevLett.102.252301}
}

@article{PhysRevLett.52.2344,
  title = {{New Scaling Phenomena in Nuclear Structure Functions}},
  author = {Dat\'e, Schin and Saito, Koichi and Sumiyoshi, Hiroyuki and Tezuka, Hirokazu},
  journal = {Phys. Rev. Lett.},
  volume = {52},
  issue = {26},
  pages = {2344--2347},
  numpages = {0},
  year = {1984},
  month = {Jun},
  publisher = {American Physical Society},
  doi = {10.1103/PhysRevLett.52.2344},
  url = {https://link.aps.org/doi/10.1103/PhysRevLett.52.2344}
}

@article{FRANKFURT1988235,
  title = {Hard nuclear processes and microscopic nuclear structure},
  author = {Leonid Frankfurt and Mark Strikman},
  journal = {Physics Reports},
  volume = {160},
  number = {5},
  pages = {235-427},
  year = {1988},
  issn = {0370-1573},
  doi = {10.1016/0370-1573(88)90179-2},
  url = {https://www.sciencedirect.com/science/article/pii/0370157388901792}
}

@article{CLAS:2019vsb,
    author = "Schmookler, B. and others",
    collaboration = "CLAS",
    title = "{Modified structure of protons and neutrons in correlated pairs}",
    doi = "10.1038/s41586-019-0925-9",
    journal = "Nature",
    volume = "566",
    number = "7744",
    pages = "354--358",
    year = "2019"
}

@article{PhysRevLett.124.092002,
  title = {{Neutron Valence Structure from Nuclear Deep Inelastic Scattering}},
  author = {Segarra, E. P. and Schmidt, A. and Kutz, T. and Higinbotham, D. W. and Piasetzky, E. and Strikman, M. and Weinstein, L. B. and Hen, O.},
  journal = {Phys. Rev. Lett.},
  volume = {124},
  issue = {9},
  pages = {092002},
  numpages = {6},
  year = {2020},
  month = {Mar},
  publisher = {American Physical Society},
  doi = {10.1103/PhysRevLett.124.092002},
  url = {https://link.aps.org/doi/10.1103/PhysRevLett.124.092002}
}

@article{Xu:2019wso,
    author = "Xu, Ji and Yuan, Feng",
    title = "{Gluonic Probe for the Short Range Correlation in Nucleus}",
    doi = "10.1016/j.physletb.2019.135187",
    journal = "Phys. Lett. B",
    volume = "801",
    pages = "135187",
    year = "2020"
}

@article{PhysRevD.104.033002,
  title = {{Nuclear effects in neutrino-nucleus DIS and a probe for short-range correlations}},
  author = {Huang, Fei and Xu, Ji and Yang, Xing-Hua},
  journal = {Phys. Rev. D},
  volume = {104},
  issue = {3},
  pages = {033002},
  numpages = {8},
  year = {2021},
  month = {Aug},
  publisher = {American Physical Society},
  doi = {10.1103/PhysRevD.104.033002},
  url = {https://link.aps.org/doi/10.1103/PhysRevD.104.033002}
}

@article{PhysRevD.108.053005,
  title = {Nuclear effects in extracting ${\mathrm{sin}}^{2}{\ensuremath{\theta}}_{\mathrm{W}}$ and a probe for short-range correlations},
  author = {Yang, Xing-Hua and Huang, Fei and Xu, Ji},
  journal = {Phys. Rev. D},
  volume = {108},
  issue = {5},
  pages = {053005},
  numpages = {9},
  year = {2023},
  month = {Sep},
  publisher = {American Physical Society},
  doi = {10.1103/PhysRevD.108.053005},
  url = {https://link.aps.org/doi/10.1103/PhysRevD.108.053005}
}

@article{Huang:2025kmd,
    author = "Huang, Fei and Hu, Shu-Man and Li, De-Min and Xu, Ji",
    title = "{Test for universality of short-range correlations in pion-induced Drell{\textendash}Yan process}",
    doi = "10.1140/epjc/s10052-025-14960-x",
    journal = "Eur. Phys. J. C",
    volume = "85",
    number = "10",
    pages = "1225",
    year = "2025"
}

@article{PhysRevD.96.114005,
  title = {Nuclear structure functions at a future electron-ion collider},
  author = {Aschenauer, E. C. and Fazio, S. and Lamont, M. A. C. and Paukkunen, H. and Zurita, P.},
  journal = {Phys. Rev. D},
  volume = {96},
  issue = {11},
  pages = {114005},
  numpages = {20},
  year = {2017},
  month = {Dec},
  publisher = {American Physical Society},
  doi = {10.1103/PhysRevD.96.114005},
  url = {https://link.aps.org/doi/10.1103/PhysRevD.96.114005}
}

@article{Eskola:2013aya,
  author = "Eskola, Kari J. and Paukkunen, Hannu and Salgado, Carlos A.",
  title = "{A perturbative QCD study of dijets in p+Pb collisions at the LHC}",
  doi = "10.1007/JHEP10(2013)213",
  journal = "JHEP",
  volume = "10",
  pages = "213",
  year = "2013"
}

@article{PhysRevD.105.L031504,
  title = {Light-nuclei gluons from dijet production in proton-oxygen collisions},
  author = {Paakkinen, Petja},
  journal = {Phys. Rev. D},
  volume = {105},
  issue = {3},
  pages = {L031504},
  numpages = {7},
  year = {2022},
  month = {Feb},
  publisher = {American Physical Society},
  doi = {10.1103/PhysRevD.105.L031504},
  url = {https://link.aps.org/doi/10.1103/PhysRevD.105.L031504}
}

@article{Eskola:2021nhw,
    author = "Eskola, Kari J. and Paakkinen, Petja and Paukkunen, Hannu and Salgado, Carlos A.",
    title = "{EPPS21: a global QCD analysis of nuclear PDFs}",
    doi = "10.1140/epjc/s10052-022-10359-0",
    journal = "Eur. Phys. J. C",
    volume = "82",
    number = "5",
    pages = "413",
    year = "2022"
}

@article{PhysRevC.76.014311,
  title = {Neutron density distributions from antiprotonic $^{208}\mathrm{Pb}$ and $^{209}\mathrm{Bi}$ atoms},
  author = {K\l{}os, B. and Trzci\ifmmode \acute{n}\else \'{n}\fi{}ska, A. and Jastrz\ifmmode \mbox{\k{e}}\else \k{e}\fi{}bski, J. and Czosnyka, T. and Kisieli\ifmmode \acute{n}\else \'{n}\fi{}ski, M. and Lubi\ifmmode \acute{n}\else \'{n}\fi{}ski, P. and Napiorkowski, P. and Pie\ifmmode \acute{n}\else \'{n}\fi{}kowski, L. and Hartmann, F. J. and Ketzer, B. and Ring, P. and Schmidt, R. and Egidy, T. von and Smola\ifmmode \acute{n}\else \'{n}\fi{}czuk, R. and Wycech, S. and Gulda, K. and Kurcewicz, W. and Widmann, E. and Brown, B. A.},
  journal = {Phys. Rev. C},
  volume = {76},
  issue = {1},
  pages = {014311},
  numpages = {13},
  year = {2007},
  month = {Jul},
  publisher = {American Physical Society},
  doi = {10.1103/PhysRevC.76.014311},
  url = {https://link.aps.org/doi/10.1103/PhysRevC.76.014311}
}

\end{document}